\documentclass[11pt,a4paper]{article} 
\pdfoutput=1
\usepackage[no-natbib-sort]{my-jheppub}

\usepackage{amsmath, amssymb}
\newcommand\numberthis{\addtocounter{equation}{1}\tag{\theequation}}
\usepackage{bm}
\usepackage{environ}
\usepackage{mathrsfs}
\usepackage{array,arydshln}

\usepackage{graphicx,epsfig}
\usepackage{epic}
\usepackage{youngtab}
\usepackage{float}
\usepackage{color}
\definecolor{maroon}{rgb}{0.8,0.3,0.}

\usepackage{slashed}
\usepackage[nodayofweek]{date time}

\usepackage{hyperref}

\usepackage{aurical}
\usepackage[T1]{fontenc}

\allowdisplaybreaks

\newcommand{\be}{\begin{equation}}
\newcommand{\ee}{\end{equation}}

\newcommand{\mc}{\mathcal }

\newcommand{\bv}{\begin{verbatim}}
\newcommand{\ev}{\end{verbatim}}

\title{Virasoro vacuum block at  next-to-leading order in the 
heavy-light limit}
\author[a]{Matteo Beccaria} 
\author[a]{, Alberto Fachechi} 
\author[a]{,  and  Guido Macorini} 

\abstract{ 
We consider the semiclassical limit of the vacuum Virasoro block describing the diagonal 4-point correlation
functions on the sphere. At large central charge $c$, after exponentiation, 
it depends on two fixed ratios $h_{\rm H}/c$ and $h_{\rm L}/c$, where $h_{\rm H, L}$ are the conformal dimensions
of the 4-point function operators. The semiclassical block may be expanded in powers of the 
light ratio $h_{\rm L}/c$
and the leading non-trivial (linear) order is known in closed form as a function of $h_{\rm H}/c$. 
Recently, this contribution has been  matched against AdS$_{3}$ gravity calculations
where  heavy operators build up a classical  geometry corresponding to a BTZ black hole, 
while the light operators are described by a geodesic in this background.
Here, we compute for the first time the next-to-leading quadratic correction $\mc O((h_{\rm L}/c)^{2})$,
again in closed form 
for generic heavy operator ratio $h_{\rm H}/c$. The result is a highly non-trivial extension of the leading order
and may be relevant for further refined AdS$_{3}$/CFT$_{2}$ tests. Applications to the 
two-interval R\'enyi entropy are also presented.
\vfill }

\affiliation[a]{Dipartimento di Matematica e Fisica Ennio De Giorgi,\\
Universit\`a del Salento \& INFN, Via Arnesano, 73100 Lecce, 
Italy}

\emailAdd{matteo.beccaria@le.infn.it} 
\emailAdd{alberto.fachechi@le.infn.it} 
\emailAdd{macorini@nbi.ku.dk}

\begin{document}



\maketitle
\flushbottom

\section{Introduction}

Conformal field theories are characterised by their operator spectrum and fusion rules. From this basic data, 
correlation functions may be computed by repeatedly applying the operator product expansion (OPE). 
This procedure is conveniently organised with respect to the representation content of the  
conformal algebra.
Its main ingredients are the global conformal blocks that are fully determined by conformal symmetry
\cite{Ferrara:1972cq,Ferrara:1971vh,Ferrara:1973vz,Ferrara:1974ny,Belavin:1984vu,Dolan:2000ut,Dolan:2003hv}.
In two dimensions, the global conformal algebra is enhanced to the Virasoro algebra 
$\rm Vir\oplus \overline{\rm Vir}$ and the associated decomposition of correlation functions
in terms of Virasoro conformal blocks is a central problem in the study of dynamical properties. 
In particular, knowledge of Virasoro conformal blocks is important for the  bootstrap program \cite{Rattazzi:2008pe},  in Liouville theory \cite{Harlow:2011ny},  AGT correspondence \cite{Alday:2009aq,Alba:2010qc}, and, of course, 
in the context of AdS$_{3}$/CFT$_{2}$ duality as we shall discuss later.

In this paper, we focus on the determination of the vacuum
Virasoro conformal blocks for the 4-point correlation functions on the sphere. Virasoro blocks
encode the contributions from the virtual exchange of the states associated with a given  (internal) 
conformal family. They are functions $\mc F(c, \bm{h}, h_{p}, z)$ that depend on the central charge $c$, 
the external and internal operators conformal weights  $\bm{h} =\{h_{i}\}_{i=1,\dots,4}$ and $h_{p}$, 
and the conformally invariant cross-ratio $z$ of the four points. 

Virasoro blocks are known in closed form only in a few special cases, typically involving degenerate fields
\cite{francesco1999conformal}.
Series expansions in  various scattering channels  may be derived by brute force computation
 exploiting the Virasoro algebra or by the recursion relations derived by 
Zamolodchikov \cite{Zamolodchikov:1985ie,Zamolodchikov:1995aa}. Recently, a remarkable progress in this
direction has been 
achieved in \cite{Perlmutter:2015iya} where three different closed-form perturbative expansions 
have been obtained for the 4-point Virasoro blocks on the sphere. 
These representations are an important computational tool to examine the semiclassical regime, {\em i.e.} a large central charge limit where $c\to\infty$ with some of the ratios
$h_{i}/c$ being fixed \cite{Litvinov:2013sxa}. Of course, one of the main motivations for considering 
this limit is that it is suitable for a holographic study by means of the AdS/CFT duality. 
Following standard usage, the operators with growing 
conformal dimension $h\sim c$ are called {\em heavy}, while operators with fixed $h$ are called {\em light}.
 An important  and well studied tractable case is the {\em heavy-light} limit 
 \cite{Fitzpatrick:2014vua} associated with the following 4-point function 
 \be
 \label{1.1}
\text{Heavy-Light:}\qquad \langle \Phi_{H}\,\Phi_{H}\,\Phi_{L}\,\Phi_{L}\rangle.
 \ee
In this case, the relevant Virasoro block is the one associated with the conformal family of the identity operator,
{\em i.e.} the so-called vacuum block. The AdS$_{3}$ bulk interpretation 
is that the heavy operators build up a classical asymptotically AdS$_{3}$ 
geometry corresponding to a BTZ black hole \cite{banados1992black}, while 
the light operators are described by a geodesic probing this background solution. The heavy-light 
case  (\ref{1.1}) may be generalised in various ways
\cite{Fitzpatrick:2014vua,Hijano:2015rla,Fitzpatrick:2015zha,Fitzpatrick:2015qma,Fitzpatrick:2015foa}.
In particular, it is interesting to regard it as a special case of the general fully-heavy situation where all ratios $h_{i}/c$
are fixed as $c\to \infty$, but two of them are taken to be small in the end. This is legitimate because it has been 
proved that there is no problem with the order of limits \cite{Fitzpatrick:2015zha}.
The perturbative expansion with respect to the small (fixed) ratios $h_{\rm L}/c$ is potentially
quite interesting.
In the bulk AdS$_{3}$ gravity theory,  it accounts for the back-reaction of the particles on the geometry and
 describes the  corrections to the emergent thermality of heavy microstates at large $c$ \cite{Fitzpatrick:2015zha}. 
Recently, this holographic computation of the semiclassical 4-point function at first order in $h_{\rm L}/c$
 has been fully clarified in \cite{Hijano:2015qja}.
To this aim, a bulk construction has been presented based on the geodesic Witten diagrams recently introduced 
in \cite{Hijano:2015zsa}, evaluated in locally AdS$_{3}$ geometries generated by the back-reaction of the 
heavy operators.

From the point of view of the 2d boundary CFT, the semiclassical limit is well defined due to the 
exponentiation property of the 
Virasoro blocks \cite{Belavin:1984vu,Zamolodchikov:1985ie,zamolodchikov1986two,Harlow:2011ny}.
Indeed, at large central charge we have 
\be
\label{1.2}
\log \mc F(c,\bm{h}, h_{p}, z) = -\frac{c}{6}\,f(\bm{\lambda}, \lambda_{p}, z)+\mc O(c^{0}), 
\ee
where $\lambda_{i}$ are the fixed ratios $h_{i}/c$. 
We now specialise to the subject of this paper, {\em i.e.}
the vacuum Virasoro conformal block that contains the contributions from the identity 
conformal family with $h_{p}=0$. 
In this case, fusion onto the identity requires to consider a pairwise correlator with  
$h_{1}=h_{2}$ and $h_{3}=h_{4}$. 
The function in the r.h.s. of (\ref{1.2}) is then  $f(\lambda_{1}, \lambda_{3}, z)$, 
symmetric under the exchange  $\lambda_{1} \leftrightarrow \lambda_{3}$. According to the above discussion, 
the interest in AdS/CFT tests suggests to consider the following 
 expansion around $\lambda_{1}=0$, with generic $\lambda_{3}$
\be
\label{1.3}
f(\lambda_{1}, \lambda_{3}, z) = 
\lambda_{1}\,f^{(1)}( \lambda_{3}, z)+
\lambda_{1}^{2}\,f^{(2)}(\lambda_{3}, z)+\cdots.
\ee
The first correction $f^{(1)}(\lambda_{3},z)$ has been computed in 
\cite{Fitzpatrick:2014vua} in closed form. The calculation is 
based on the perturbative solution of a monodromy problem \cite{Hartman:2013mia,Harlow:2011ny} (see
also the more recent works \cite{Alkalaev:2014sma,Alkalaev:2015wia,Alkalaev:2015lca}). \footnote{
In the context of holographic calculations of entanglement entropy \cite{Hartman:2013mia,Faulkner:2013yia}, the monodromy approach  is related to the problem of finding  solutions of Einstein equations with 
given boundary behaviour.}
The holographic setup in \cite{Hijano:2015qja} is indeed able to reproduce $f^{(1)}(\lambda_{3},z)$
 by means of a bulk gravity  calculation. In order to sharpen both this picture and the associated tools, 
we believe that it is important to extend the expansion (\ref{1.3}) including  
 the much more difficult next order contribution 
$f^{(2)}(\lambda_{3}, z)$. In principle, this task can be accomplished by solving  the 
monodromy problem at higher order, but technical difficulties must be overcome.
The aim of this paper is precisely that of providing the exact expression for $f^{(2)}(\lambda_{3}, z)$.
Although our derivation will be somewhat 
heuristic, the  final result -- presented in (\ref{4.1})  -- passes several non-trivial tests. In particular, it
agrees with the explicit calculation of the perturbative
Virasoro vacuum block at large orders in the $s$-channel, that we have computed 
exploiting the powerful recent 
results of \cite{Perlmutter:2015iya}.

The plan of the paper is the following. In Sec.~\ref{sec:notation} we set up the notation. In Sec.~\ref{sec:semiclassical}
we discuss the semiclassical limit of the vacuum block in some details. We include explicit results from the high order
solution of the recursion relations obeyed by the block. This allows to derive a series expansion of the 
(perturbative) exponentiated block. In Sec.~\ref{sec:exact} we present our main results. 
Finally, in Sec.~\ref{sec:entropy} we give a simple 
application to the study of two-interval mutual R\'enyi information.
The details of our derivation and some additional exact results are collected in App.~\ref{app:details}.
Other appendices collect long expressions that may be useful for the reader.

\section{Notation and preliminary considerations}
\label{sec:notation}

The general form of the OPE between two Virasoro primary fields $\Phi_{1}$, $\Phi_{2}$
with conformal dimensions $h_{1}$, $h_{2}$ is  
\be
\label{2.1}
\Phi_{1}(z, \overline z)\,\Phi_{2}(0,0) = \sum_{p} C^{p}_{12}\,|z|^{2\,(h_{p}-h_{1}-h_{2})}\,\Psi_{p}
(z, \overline z\,|\,0,0),
\ee
where we sum over all primaries $\Phi_{p}(z,\overline z)$ and where $\Psi_{p}(z, \overline z\,|\,0,0)$ takes into account the contributions from the conformal family of $\Phi_{p}(z,\overline z)$, {\em i.e.} the primary and its descendants.
The explicit expression of $\Psi_{p}(z, \overline z\,|\,0,0)$ involves a sum over descendants and may be completely fixed
by using the Virasoro algebra  (see \cite{FigueroaO'Farrill:1990dy} for efficient algorithms).
The 3-point couplings $C^{p}_{12}$ are among the basic CFT data.  A 4-point correlation function of primaries has 
indeed the form 
\be
\label{2.2}
\langle\Phi_{1}(\infty)\,\Phi_{2}(1)\,\Phi_{3}(z)\,\Phi_{4}(0)\rangle = \frac{1}{|z|^{2\,(h_{3}+h_{4})}}
\sum_{p} C_{12,p}\,C^{p}_{34}\,|\mc F(c, \bm{h}, h_{p}, z)|^{2},
\ee
where we used global conformal invariance to fix three operators at $\infty, 1, 0$. The letter $\bm{h}$ stands 
collectively for the conformal dimensions
$h_{1}, \dots, h_{4}$, and the index in the fusion coefficients is raised by the Zamolodchikov metric
$g_{ab} = \langle\Phi_{a}\,|\,\Phi_{b}\rangle$. The Virasoro block associated with the internal primary $\Phi_{p}(z,\overline z)$ is the function $\mc F(c, \bm{h}, h_{p}, z)$ in (\ref{2.2}). It may be expanded in powers of $z$, {\em i.e.}
in the Virasoro level $\ell$
\be
\label{2.3}
\mc F(c, \bm{h}, h_{p}, z) = z^{h_{p}}\,\sum_{\ell=0}^{\infty} \mc F_{\ell}(c, \bm{h}, h_{p})\,z^{\ell}.
\ee
This expression may be further refined by separating the contributions of the level $q$ quasi-primaries
with respect to the global conformal algebra $\mathfrak{sl}(2, \mathbb R)\oplus \mathfrak{sl}(2, \mathbb R)$,
see \cite{Perlmutter:2015iya} for a clean presentation. The result is ($h_{ij} \equiv h_{i}-h_{j}$)
\be
\label{2.4}
\mc F(c,\bm{h}, h_{p}, z) = z^{h_{p}}\,\sum_{q=0}^{\infty}\chi_{q}(c, \bm{h}, h_{p})\,
z^{q}\,{}_{2}F_{1}(h_{p}+q+h_{12}, h_{p}+q+h_{34}, 2\,(h_{p}+q) ; z).
\ee
Here, $\chi_{q}(c, \bm{h}, h_{p})$ is a well-definite sum over the level $q$ quasi-primaries
and it involves the $1\times 2\to q$, $3\times 4\to q$ fusion coefficients. It is completely fixed by Virasoro symmetry.
In all cases $\chi_{0}=1$ and $\chi_{1}=0$. For the vacuum block, obtained with $h_{p}\to 0$, we need 
$h_{12}=h_{34}=0$. Setting now $\bm{h}=\{h_{1}, h_{3}\}$,  (\ref{2.4}) specialises to 
\be
\label{2.5}
\mc F_{\rm vac}(c, \bm{h}, z) = \mathop{\sum_{q=0}}_{q\,\rm even}^{\infty}\chi_{{\rm vac},q}(c, \bm{h})\,z^{q}\,
{}_{2}F_{1}(q,q,2q; z).
\ee
The first terms are rather simple
\be
\label{2.6}
\chi_{{\rm vac},2}(c,\bm{h}) = \frac{2\,h_{1}\,h_{3}}{c}, \qquad
\chi_{{\rm vac},4}(c,\bm{h}) = \frac{10\,(h_{1}^{2}+\tfrac{h_{1}}{5}),(h_{3}^{2}+\tfrac{h_{3}}{5})}
{c\,(5\,c+22)},
\ee
and, for larger $q$, they are always rational functions of $c$, $h_{1}$, $h_{3}$ with rational numerical coefficients.
Using the method of \cite{Perlmutter:2015iya} to solve the $h_{p}\to 0$ limit of the recursion relation 
\be
\label{2.7}
\begin{split}
\mc F(c, \bm{h}, h_{p}, z) &= z^{h_{p}}\,_{2}F_{1}(h_{p}+h_{12}, h_{p}+h_{34}, 2h_{p}; z)\\
&+\sum_{m\ge 1, n\ge 2}^{\infty}\frac{R_{mn}(\bm{h}, h_{p})}{c-c_{mn}(h_{p})}\,
\mc F(c_{mn}(h_{p}), \bm{h}, h_{p}+mn, z),
\end{split}
\ee
we obtained the explicit rational functions $\chi_{{\rm vac},q}(c,\bm{h})$ up to $q=16$. \footnote{
The expressions are available on request. Notice that they require some tricky implementation because a naive 
coding of the 
relations in  \cite{Perlmutter:2015iya} has a complexity that grows prohibitively on  symbolic manipulation softwares.
We warn the reader that the internal sum and product in (2.28) of \cite{Perlmutter:2015iya} must be swapped.
} Notice that from the elementary relation 
\be
\label{2.8}
_{2}F_{1}\bigg(a,c-b,c;\frac{z}{z-1}\bigg) = (1-z)^{a}\,_{2}F_{1}(a,b,c; z),
\ee
we check  invariance of (\ref{2.5}) under the tranformation $z\to\tfrac{z}{z-1}$,
\be
\label{2.9}
\mc F_{\rm vac}(c, \bm{h}, z) = \mc F_{\rm vac}\bigg(c, \bm{h}, \frac{z}{z-1}\bigg),
\ee
that is nothing but the (bootstrap) symmetry between the $s$ and $t$ scattering channels.

\section{Semiclassical limit}
\label{sec:semiclassical}

As discussed in the introduction, the semiclassical limit of the vacuum block is defined as 
\be
\label{3.1}
c\to \infty, \quad \lambda_{1}=\frac{h_{1}}{c}, \ \ \lambda_{3}=\frac{h_{3}}{c}\ \ \text{fixed}.
\ee
In this limit, there are arguments 
\cite{Belavin:1984vu,Zamolodchikov:1985ie,zamolodchikov1986two,Harlow:2011ny}
to expect that the block exponentiates according to the relation ($\bm{\lambda} = \{\lambda_{1}, \lambda_{3}\}$)
\be
\label{3.2}
\log\mc F(c, \bm{h}, z) = -\frac{c}{6}\,f(\bm{\lambda}, z)+\mc O(c^{0}).
\ee
The function $f(\bm{\lambda}, z)$ is not known in closed form, but it may be considered order by order 
at small values of one of the ratios $\bm\lambda$. We define the coefficient functions 
$f^{(\ell)}(\lambda, z)$ by expanding in powers of $\lambda_{1}$
\be
\label{3.3}
f(\bm{\lambda},  z) =  \sum_{\ell_{1}=1}^{\infty}
\lambda_{1}^{\ell_{1}}\,f^{(\ell_{1})}(\lambda_{3}, z).
\ee
The exact expression of the first function $f^{(1)}(\lambda, z)$ appearing in (\ref{3.3}) is known 
from the calculation in \cite{Fitzpatrick:2014vua} and reads
\be
\label{3.4}
\begin{split}
f^{(1)}(\lambda, z) &= 12\,\log\bigg(\frac{1-(1-z)^{\alpha}}{\alpha\,z}\bigg)+6\,(1-\alpha)\,\log(1-z),\\
\alpha &= \sqrt{1-24\,\lambda}.
\end{split} 
\ee
In the following, it will be important to further refine the expansion (\ref{3.3}) by separating out 
different 
powers of the ratio $\lambda_{3}$, as follows
\be
\label{3.5}
f^{(\ell_{1})}(\lambda_{3}, z) = \sum_{\ell_{3}=1}^{\infty}
\lambda_{3}^{\ell_{3}}\,f^{(\ell_{1},\ell_{3})}(z),
\ee
where we have the obvious symmetry $f^{(\ell,\ell')}(z) = f^{(\ell',\ell)}(z)$. In particular,
from the result (\ref{3.4}), we may obtain all the functions $f^{(1, \ell)}(z)$. The first cases are
\be
\label{3.6}
\begin{split}
& f^{(1,1)}(z) = -12\,U_{2}, \\
& f^{(1,2)}(z) = 12\,U_{2}^{2}+\frac{72}{5}\,U_{3}, \\
& f^{(1,3)}(z) = -16\,U_{2}^{3}-\frac{168}{5}\,U_{2}\,U_{3}-\frac{648}{35}\,U_{4}, \\
& \cdots \\
& \text{with}\   U_{q}(z) = z^{q}\,\log^{q-2}(1-z)\,_{2}F_{1}(q,q,2q; z).
\end{split}
\ee
The function   $f^{(2)}(\lambda, z)$ in (\ref{3.3}) is not known and is the next-to-leading correction when one of the 
ratios $\bm\lambda$ is small. The main result of this paper is a closed form for this contribution, analogous
to (\ref{3.4}).

\subsection{Exact expansions from the recursion relations}

Using our explicit data for $\chi_{{\rm vac},q}(c,\bm{h})$, we checked that the exponentiation property
written in (\ref{3.2})
holds perturbatively in small $z$ at least up to the order $\mc O(z^{17})$. To clarify what one gets, we 
write here the first terms of the function $f(\bm{\lambda}, z)$ \footnote{All other terms 
are again available on request, although they can be computed directly by evaluating the limit (\ref{3.1}).}
\be
\label{3.7}
\begin{split}
 f(\bm{\lambda}, & z) =-12 \left(\lambda _1 \lambda _3\right) z^2-12 \left(\lambda _1 \lambda _3\right)
   z^3+\left(\frac{264}{5} \lambda _3^2 \lambda _1^2-\frac{12}{5} \lambda _3 \lambda
   _1^2-\frac{12}{5} \lambda _3^2 \lambda _1-\frac{54 \lambda _3 \lambda _1}{5}\right)
   z^4\\
   &+\left(\frac{528}{5} \lambda _3^2 \lambda _1^2-\frac{24}{5} \lambda _3 \lambda
   _1^2-\frac{24}{5} \lambda _3^2 \lambda _1-\frac{48 \lambda _3 \lambda _1}{5}\right)
   z^5\\
   &+\left(-\frac{24064}{35} \lambda _3^3 \lambda _1^3+\frac{1776}{35} \lambda _3^2
   \lambda _1^3-\frac{32}{35} \lambda _3 \lambda _1^3+\frac{1776}{35} \lambda _3^3 \lambda
   _1^2+\frac{1014}{7} \lambda _3^2 \lambda _1^2-\frac{234}{35} \lambda _3 \lambda
   _1^2\right. \\
   & \left.  -\frac{32}{35} \lambda _3^3 \lambda _1-\frac{234}{35} \lambda _3^2 \lambda
   _1-\frac{60 \lambda _3 \lambda _1}{7}\right) z^6+\left(-\frac{72192}{35} \lambda _3^3
   \lambda _1^3+\frac{5328}{35} \lambda _3^2 \lambda _1^3-\frac{96}{35} \lambda _3 \lambda
   _1^3\right. \\
   & \left. +\frac{5328}{35} \lambda _3^3 \lambda _1^2+\frac{1194}{7} \lambda _3^2 \lambda
   _1^2-\frac{282}{35} \lambda _3 \lambda _1^2-\frac{96}{35} \lambda _3^3 \lambda
   _1-\frac{282}{35} \lambda _3^2 \lambda _1-\frac{54 \lambda _3 \lambda _1}{7}\right)
   z^7+\cdots\ .
\end{split}
\ee
We can collect the various powers of $\bm\lambda$ and extract series expansions for the functions
$f^{(\ell,\ell')}(z)$ defined in (\ref{3.5}). The first instances of the first order 
contributions $f^{(1,\ell)}(z)$ are 
\be
\label{3.8}
\begin{split}
f^{(1,1)}(z) &= -12 z^2-12 z^3-\frac{54 z^4}{5}-\frac{48 z^5}{5}-\frac{60 z^6}{7}-\frac{54 z^7}{7}-7
   z^8-\frac{32 z^9}{5}-\frac{324 z^{10}}{55}+\dots, \\
f^{(1,2)}(z) &= -\frac{12 z^4}{5}-\frac{24 z^5}{5}-\frac{234 z^6}{35}-\frac{282 z^7}{35}-\frac{1578
   z^8}{175}-\frac{1692 z^9}{175}-\frac{3886 z^{10}}{385}-\frac{3986
   z^{11}}{385}+\dots, \\
f^{(1,3)}(z) &= -\frac{32 z^6}{35}-\frac{96 z^7}{35}-\frac{908 z^8}{175}-\frac{1392 z^9}{175}-\frac{20826
   z^{10}}{1925}-\frac{150 z^{11}}{11}-\frac{14293984 z^{12}}{875875}+\dots \ .
\end{split}
\ee
One can check immediately that the expansions in (\ref{3.8}) agree with the formulas in (\ref{3.6}), as they 
should. Going further, from (\ref{3.7}) -- and including additional terms --   we can write down 
the first cases of the  second order 
contributions $f^{(2,\ell)}(z)$. They are 
\be
\label{3.9}
\begin{split}
f^{(2,2)}(z) &= \frac{264 z^4}{5}+\frac{528 z^5}{5}+\frac{1014 z^6}{7}+\frac{1194 z^7}{7}+\frac{162516
   z^8}{875}+\frac{169164 z^9}{875}+\frac{376744 z^{10}}{1925}\\
   &+\frac{374564
   z^{11}}{1925}+\frac{1171852056 z^{12}}{6131125}+\frac{1141630738
   z^{13}}{6131125}+\frac{57446497 z^{14}}{318500}\\
   &+\frac{121933377
   z^{15}}{700700}+\frac{4487292704 z^{16}}{26801775}+\frac{439701599
   z^{17}}{2734875}+\dots,
 \end{split}
 \ee
 and 
 \be
 \label{3.10}
 \begin{split}
f^{(2,3)}(z) &= \frac{1776 z^6}{35}+\frac{5328 z^7}{35}+\frac{250356 z^8}{875}+\frac{379824
   z^9}{875}+\frac{801346 z^{10}}{1375}+\frac{1393606 z^{11}}{1925}\\
   &+\frac{26149378198
   z^{12}}{30655625}+\frac{29672482498 z^{13}}{30655625}+\frac{65496704213
   z^{14}}{61311250}+\frac{2022337201 z^{15}}{1751750}\\
   &+\frac{3838174474811
   z^{16}}{3126873750}+\frac{2014459667366 z^{17}}{1563436875}+\dots.
\end{split}
\ee
We have written many terms in (\ref{3.9}) and (\ref{3.10}) to emphasise that it is non trivial to identify their resummation
in terms of a closed function of $z$ as it was possible in 
(\ref{3.6}). The solution will be presented in the next section
and indeed will be much more involved than the leading order (\ref{3.6}).
Here, we just 
notice that the relevant terms from (\ref{3.7}) are
\be
\label{3.11}
\begin{split}
f^{(2)}&(\lambda, z) = \left(\frac{264 \lambda ^2}{5}-\frac{12 \lambda }{5}\right) z^4+\left(\frac{528 \lambda
   ^2}{5}-\frac{24 \lambda }{5}\right) z^5+\left(\frac{1776 \lambda ^3}{35}+\frac{1014
   \lambda ^2}{7}-\frac{234 \lambda }{35}\right) z^6\\
   &+\left(\frac{5328 \lambda
   ^3}{35}+\frac{1194 \lambda ^2}{7}-\frac{282 \lambda }{35}\right) z^7+\left(\frac{36144
   \lambda ^4}{875}+\frac{250356 \lambda ^3}{875}+\frac{162516 \lambda ^2}{875}-\frac{1578
   \lambda }{175}\right) z^8\\
   &+\left(\frac{144576 \lambda ^4}{875}+\frac{379824 \lambda
   ^3}{875}+\frac{169164 \lambda ^2}{875}-\frac{1692 \lambda }{175}\right)
   z^9\\
   &+\left(\frac{301248 \lambda ^5}{9625}+\frac{541336 \lambda ^4}{1375}+\frac{801346
   \lambda ^3}{1375}+\frac{376744 \lambda ^2}{1925}-\frac{3886 \lambda }{385}\right)
   z^{10}\\
   &+\left(\frac{301248 \lambda ^5}{1925}+\frac{1403848 \lambda
   ^4}{1925}+\frac{1393606 \lambda ^3}{1925}+\frac{374564 \lambda ^2}{1925}-\frac{3986
   \lambda }{385}\right) z^{11}+\dots.
 \end{split}
\ee
These may be used as a check, although we shall also provide  much longer expansions at specialised values of 
$\lambda$ to further check. Of course, the terms linear in $\lambda$ in (\ref{3.11}) define the function 
$f^{(2,1)}(z)$. This is equal to  $f^{(1,2)}(z)$ and indeed the expansion is the same as that in the second line
of (\ref{3.8}). In the following, it will be convenient to trade $\lambda$ for the same $\alpha$ parameters as in (\ref{3.4}),
{\em i.e.} $\alpha = \sqrt{1-24\,\lambda}$. With a little abuse of notation, we shall denote $f^{(2)}(\lambda(\alpha), z)
\to f^{(2)}(\alpha, z)$ since no confusion should arise. The first terms of (\ref{3.11}) after this redefinition 
are then 
\be
\label{3.12}
\begin{split}
f^{(2)}&(\alpha, z) = \frac{1}{120} (\alpha^{2} -1)\left(11 \alpha ^2+1\right) z^4+\frac{1}{60} 
   (\alpha^{2}-1) \left(11 \alpha ^2+1\right) z^5\\
   &-\frac{(\alpha^{2} -1) \left(37
   \alpha ^4-2609 \alpha ^2-236\right)}{10080}\,z^{6}-\frac{(\alpha^{2} -1)
   \left(37 \alpha ^4-1069 \alpha ^2-96\right)}{3360}\,z^{7}\\
   &+\frac{(\alpha^{2} -1)
   \left(251 \alpha ^6-42479 \alpha ^4+734269 \alpha ^2+65399\right)
   }{2016000}\,z^{8}+\dots\ .
\end{split}
\ee
Many additional terms are collected in App.~\ref{app:f2}.

\newpage
\section{The exact expression for the NLO contribution $f^{(2)}(\lambda, z)$}
\label{sec:exact}

Our main result is the following closed expression for the derivative $\partial_{z}f^{(2)}(\lambda, z)$.
This is what is obtained from the solution of the second order monodromy problem according to the 
procedure discussed in App.~\ref{app:details}.
\be
\label{4.1}
\begin{split}
\partial_{z}\,f^{(2)}(\lambda, z) = \rm (I)+(II)+(III), 
\end{split}
\ee
with
\be
\label{4.2}
\begin{split}
{\rm (I)} &= \frac{36}{{(1-z) z \left(1-(1-z)^{\alpha }\right)^3}} \bigg[
 (\alpha +1)\, z\\
 & \qquad - (1-z)^{\alpha } (\alpha\,  z+4\,   \alpha\,  z\, \log (1-z)-8 \alpha\,  z\, \log z-9 \,z+6)\\
 &\qquad + (1-z)^{2 \alpha } (-\alpha  \,z-4 \,\alpha\, 
 z\, \log (1-z)+8\, \alpha\,  z\, \log z+3\, z+8)\\
& \qquad +(1-z)^{3 \alpha }\, ((\alpha +3)\, z-2) 
 \bigg],
\end{split}
\ee 
\be
\label{4.3}
\begin{split}
{\rm (II)} &= \frac{144 \alpha  \left((1-z)^{\alpha }+1\right) (1-z)^{\alpha -1} (\pi  \cot (\pi  \alpha )+2
   \psi ^{(0)}(\alpha )+2 \gamma )}{\left(1-(1-z)^{\alpha }\right)^3}
\end{split}
\ee 
\be
\begin{split}
\label{4.4}
{\rm (III)} &= -\frac{72 \left((1-z)^{\alpha }+1\right)^2}{(1-z)
   \left(1-(1-z)^{\alpha }\right)^3}\, \, _2F_1(1,-\alpha ,1-\alpha ;1-z)\\
   &-\frac{72 \left((1-z)^{\alpha }+1\right)}{(\alpha -1) z \left(1-(1-z)^{\alpha }\right)^2}\,
    \, _2F_1(1,-\alpha ,2-\alpha ;1-z)\\
   &+\frac{\left(-144 (\alpha
   -1) (1-z) \left((z-1)^2\right)^{\alpha }+144 (1-z)^{2 \alpha } \left(\alpha +\alpha  z
   (1-z)^{\alpha }+z-1\right)\right) }{(\alpha +1) z
   \left(1-(1-z)^{\alpha }\right)^3}\,\times \\
   &\qquad\qquad  _2F_1(1,\alpha +1,\alpha +2;1-z).
\end{split}
\ee 
If we expand this expression in powers of $z$ and integrate term by term -- with no additional integration constant --
we recover indeed (\ref{3.12}), including many additional non-trivial terms that may be found in 
App.~\ref{app:f2}.

\subsection{Exact expression for $f^{(2,2)}(z)$ and $f^{(2,3)}(z)$}

In order to obtain all the functions $f^{(2,\ell)}(z)$ from (\ref{4.1}), we simply have to replace $\alpha=\sqrt{1-24\,\lambda}$ and expand around $\lambda=0$. The only non-trivial piece is (III) in  (\ref{4.4}). This may be 
treated by using the algorithms described in  \cite{Huber:2005yg,Huber:2007dx}. We need the expansion of the hypergeometric functions in (\ref{4.4}) around $\alpha=1$. Setting $\alpha=1+\varepsilon$, these read
\be
\begin{split}
_{2}F_{1}(1,-\alpha,1-\alpha,z) &= 
\frac{z}{\varepsilon }+z+z \log (1-z)+1+z\,   (\log (1-z)-\text{Li}_2(z))\,\varepsilon\\
&-z \, (\text{Li}_2(z)+\text{Li}_3(z))\,\varepsilon ^2-z\,
   (\text{Li}_3(z)+\text{Li}_4(z))\,\varepsilon^{3}+\dots, \\
_{2}F_{1}(1,-\alpha,2-\alpha,z) &= 
1-z+  (-z-z \,\log (1-z)+\log (1-z))\,\varepsilon\\
&-(z-1) \, (\log
   (1-z)-\text{Li}_2(z))\,\varepsilon ^2+(z-1) \,
   (\text{Li}_2(z)+\text{Li}_3(z))\,\varepsilon ^3+\dots, \\
_{2}F_{1}(1,1+\alpha,2+\alpha,z) &= -\frac{2 (z+\log (1-z))}{z^2}
+\frac{ (-2 \,\text{Li}_2(z)+z-\log
   (1-z))}{z^2}\,\varepsilon\\
   &-\frac{ (\text{Li}_2(z)-2
   \text{Li}_3(z)+z)}{z^2}\,\varepsilon^{2}+\frac{ (\text{Li}_3(z)-2
   \text{Li}_4(z)+z)}{z^2}\,\varepsilon ^3+\dots.
   \end{split}
\ee
Using these results, we obtain 
\be
\label{4.5}
\begin{split}
f^{(2,2)}(z) &= \frac{864}{z^{3}}\bigg[ 30 z^3+\log (1-z) \left(4 (z-1) \log (1-z) (3 z-6 z \log z \right. \\
&\left. +(4
   z-6) \log (1-z))-3 (z-2) z^2\right)\bigg]\\
   &-\frac{41472 (z-1) (-\text{Li}_3(1-z)+\text{Li}_2(1-z) \log (1-z)+\zeta_{3})}{z^2}.
\end{split}
\ee
Expansion in powers of $z$ reproduces the result in (\ref{3.9}). Besides, the expression in (\ref{4.5})
is crossing invariant, see (\ref{2.9}). With more work, one can also derive the expression of $f^{(2,3)}(z)$, the main 
difficulty being the final integration of the expansion of (\ref{4.1}). The result is 
\be
\label{4.6}
\begin{split}
f^{(2,3)}(z) &= \frac{10368}{z^{4}}\bigg[2 \left(23+12 \pi ^2\right) z^4-3 z^3 \log (1-z) (z+48 z \log
   (z)-2)\\
   &-12 (z-1) z^2 \log ^2(1-z) (2 \log (z)-1)+(z-1) (11 (z-4) z+36) \log
   ^4(1-z)\\
   &-2 (z-1) z \log ^3(1-z) (-9 z+12 (z-2) \log (z)+14)\bigg]\\
   &-\frac{497664}{z^{3}}\bigg[3 z^3 \text{Li}_2(z)+\text{Li}_2(1-z) \left(3 z^3+(z-1) \log
   (1-z) (z+(z-2) \log (1-z))\right)\\
   &-(z-1) (z+(z-2) \log (1-z))
   (\text{Li}_3(1-z)-\zeta_{3})\bigg].
\end{split}
\ee
Again, expansion in powers of $z$ reproduces the result in (\ref{3.10}) and  one checks that 
(\ref{4.6}) is crossing invariant, see (\ref{2.9}).

\section{A simple application: two interval R\'enyi entropy}
\label{sec:entropy}

As a simple application, we follow the discussion in \cite{Perlmutter:2015iya} and 
connect our results for the vacuum block
with the two-interval R\'enyi entropy. This is the quantity $S_{n}$ obtained as the 
4-point correlation functions of twist fields $\Phi_{\pm}$
\be
\label{5.1}
S_{n}(z) = \frac{1}{1-n}\,\log\langle\Phi_{+}(\infty)\,\Phi_{-}(1)\,\Phi_{+}(z)\,\Phi_{-}(0)\rangle_{\mathscr C^{n}/\mathbb Z_{n}}, \qquad h_{\Phi_{\pm}} = \frac{n\,c}{24}\bigg(1-\frac{1}{n^{2}}\bigg),
\ee
where $\mathscr C^{n}/\mathbb Z_{n}$ is a cyclic orbifold with central charge $n\,c$. 
At large $c$ the twist fields behave as {\em heavy operators}. Using the exponentiation (\ref{3.2}) -- taking into
account the antiholomorphic part -- the vacuum contribution to $S_{n}$ may be shown to be simply
 \cite{Hartman:2013mia}
\be
\label{5.2}
S_{n, \rm vac}(z) = -\frac{n\,c}{3\,(1-n)}\,f\bigg(\frac{h_{\Phi}}{c}, \frac{h_{\Phi}}{c}, z\bigg),
\ee
up to subleading corrections as $c\to \infty$. Replacing (\ref{3.7}) into (\ref{5.2}) we 
obtain the function  $S_{n, \rm vac}(z)$ at order $\mc O(z^{17})$. The first 14 terms are shown is  in App.~\ref{app:renyi}. 
Of course, we reproduce the $\mc O(z^{9})$
expansion in (3.15) of \cite{Chen:2013dxa}, with generic $n$. \footnote{The quantity $I_{n}$ in \cite{Chen:2013dxa} is the mutual information equal to $-S_{n}$
plus a $\sim \log z$ term that is absent here due to our normalization, as discussed before.}

According to the analysis of  \cite{Perlmutter:2015iya}, it is convenient to organise $S_{n}$
at fixed $z$ in an expansion around $\delta n=0$, where $n=1+\delta n$. Then, 
the expansions (\ref{3.3}) and (\ref{3.5}) resum an infinite number of terms. To this aim, we define the functions 
$S^{(k)}(z)$ by writing
\be
\label{5.3}
\begin{split}
\left. S_{n, \rm vac}(z)\right|_{\mc O(c)} &= \sum_{k=1}^{\infty} S^{(k)}(z)\,(\delta n)^{k}.
\end{split}
\ee
The small $z$ expansion of the first five functions reads
\begin{align}
\label{5.4}
S^{(1)}(z) &= -\frac{z^2}{36}-\frac{z^3}{36}-\frac{z^4}{40}-\frac{z^5}{45}-\frac{5
   z^6}{252}-\frac{z^7}{56}-\frac{7 z^8}{432}+\dots,\notag \\
S^{(2)}(z) &= \frac{z^2}{18}+\frac{z^3}{18}+\frac{53 z^4}{1080}+\frac{23
   z^5}{540}+\frac{187 z^6}{5040}+\frac{493 z^7}{15120}+\frac{81
   z^8}{2800}+\dots, \notag \\
S^{(3)}(z) &= -\frac{13 z^2}{144}-\frac{13 z^3}{144}-\frac{25 z^4}{324}-\frac{83
   z^5}{1296}-\frac{8267 z^6}{155520}-\frac{769 z^7}{17280}-\frac{8009
   z^8}{212625}+\dots,\\
S^{(4)}(z) &= \frac{19 z^2}{144}+\frac{19 z^3}{144}+\frac{277 z^4}{2592}+\frac{53
   z^5}{648}+\frac{9581 z^6}{155520}+\frac{2401 z^7}{51840}+\frac{316961
   z^8}{9072000}+\dots,\notag \\
S^{(5)}(z) &= -\frac{13 z^2}{72}-\frac{13 z^3}{72}-\frac{871 z^4}{6480}-\frac{143
   z^5}{1620}-\frac{147523 z^6}{2799360}-\frac{25843
   z^7}{933120}-\frac{535277 z^8}{48988800}+\dots . \notag
\end{align}
The first four cases may be given in closed form using using our new results because
\be
\label{5.5}
\begin{split}
S^{(1)}(z) &= \frac{1}{432}\, f^{(1,1)}, \qquad S^{(2)}(z) = \frac{1}{2592}\,(f^{(1,2)}-12\,f^{(1,1)}), \\
S^{(3)}(z) &= \frac{1}{62208}(468 f^{(1,1)}-84 f^{(1,2)}+2
   f^{(1,3)}+f^{(2,2)}), \\
S^{(4)}(z) &= \frac{1}{373248}(-4104 f^{(1,1)}+1188 f^{(1,2)}-60
   f^{(1,3)}+f^{(1,4)}-30 f^{(2,2)}+f^{(2,3)}),
\end{split}
\ee
and all relevant $f^{(\ell, \ell')}(z)$ have been computed.
In particular, the terms $S^{(3)}$ and $S^{(4)}$ are an extension of the previous results since they
involve $f^{(2,2)}$ and $f^{(2,3)}$.  For completeness, we give 
the explicit expressions of the combinations in (\ref{5.5})
\be
\label{5.6}
\begin{split}
S^{(1)} &= -\frac{(z-2) \log (1-z)-2 z}{6 z}, \\
S^{(2)} &= \frac{\log (1-z) ((z-2) z+2 (z-1) \log (1-z))}{6 z^2}, \\
S^{(3)} &= \frac{1}{18 z^3}(12 (z-1) z \text{Li}_3(1-z)-12 (z-1) z \text{Li}_2(1-z) \log
   (1-z)+z^3-3 z^3 \log (1-z)\\
   &-12 z^2 \zeta_3+6 z^2 \log ^3(1-z)-12 z^2
   \log ^2(1-z)-6 z^2 \log ^2(1-z) \log (z)+6 z^2 \log (1-z)\\
   &+12 z \zeta_{3}-16 z \log ^3(1-z)+10 \log ^3(1-z)+12 z \log ^2(1-z)+6 z \log
   ^2(1-z) \log (z)),\\
S^{(4)} &= -\frac{1}{18 z^4}(72 z^4 \text{Li}_2(z)+36 z^3 \text{Li}_3(1-z)-24 z^3
   \text{Li}_3(1-z) \log (1-z)-36 z^2 \text{Li}_3(1-z)\\
   &+72 z^2
   \text{Li}_3(1-z) \log (1-z)+12 z \text{Li}_2(1-z) \left(6 z^3+2
   \left(z^2-3 z+2\right) \log ^2(1-z) \right. \\
   &\left. -3 (z-1) z \log (1-z)\right)-48 z
   \text{Li}_3(1-z) \log (1-z)-12 \pi ^2 z^4-3 z^4 \log (1-z)\\
   &+72 z^4 \log
   (1-z) \log (z)-36 z^3 \zeta_3+24 z^3 \zeta_3 \log (1-z)-6 z^3 \log
   ^4(1-z)\\
   &+18 z^3 \log ^3(1-z)+12 z^3 \log ^3(1-z) \log (z)-18 z^3 \log
   ^2(1-z)-18 z^3 \log ^2(1-z) \log (z)\\
   &+6 z^3 \log (1-z)+36 z^2 \zeta_{3}
   -72 z^2 \zeta_3 \log (1-z)+31 z^2 \log ^4(1-z)-48 z^2 \log
   ^3(1-z)\\
   &-36 z^2 \log ^3(1-z) \log (z)+18 z^2 \log ^2(1-z)+18 z^2 \log
   ^2(1-z) \log (z)+48 z \zeta_3 \log (1-z)\\
   &-46 z \log ^4(1-z)+21 \log
   ^4(1-z)+30 z \log ^3(1-z)+24 z \log ^3(1-z) \log (z)).
\end{split}
\ee
Of course, their expansions at small $z$ agree with the first four lines in (\ref{5.4}).
We conclude this section with a curious remark. The definition (\ref{5.1}) of the R\'enyi entropy requires $n$ to be integer.
Analytic continuation is possible at least at the level of the perturbative expansion in $z$, since all coefficients are
rational functions of $n$, see App.~\ref{app:renyi}. At the special value $n=1/2$, we found that it is possible to resum our
long expansion giving the closed form 
\be
\begin{split}
\left. S_{\tfrac{1}{2}, \rm vac}(z)\right|_{\mc O(c)} &= \frac{1}{6} \log \left(\frac{16}{27 (1-z)}\right)-\frac{1}{3} \sinh
   ^{-1}\left(\frac{2 z-1}{\sqrt{3}}\right)\\
   &+\frac{2}{3} \tanh
   ^{-1}\left(\frac{2 z-1}{4 \sqrt{1-z\,(1-z)}+3}\right).
\end{split}
\ee

\section{Conclusions}

In this paper we have improved our knowledge of  the semiclassical heavy-light limit by computing
the next-to-leading order corrections to the vacuum Virasoro block. Generally speaking, 
these corrections have an interesting dual interpretation at the level of the background geometry. Our explicit 
results may be discouraging due to their complexity, but could be  tested in some special limit. 
A natural extension of this work is to CFTs with extended $\mc W$-symmetry, see \cite{deBoer:2014sna}.
This is a non-trivial task because of the dependence on the additional $\mc W$-charges besides Virasoro conformal
weights. Another generalization is to consider blocks with a non-trivial  intermediate field with $h_{p}\neq 0$.
This may be possible by exploiting the recent results in \cite{Hegde:2015dqh}
about the evaluation of Wilson lines in 3D higher spin gravity. The idea is that if the chiral Wilson 
line carries a spin-$s$ charge, we can extract its part linear in this charge. 
This term must be proportional to the Virasoro block for the spin-$s$ current exchange.
Finally, we believe that it could be of some
interest to provide a cleaner derivation of the perturbative solution of the monodromy equations, possibly
exploiting its relation with the gravity equations of motion in Chern-Simons form.

\section*{Acknowledgments}
We  thank   E. Perlmutter for   useful comments.

\appendix

\section{Long expansion for the function $f^{(2)}(\alpha, z)$}
\label{app:f2}

Let us write the expansion (\ref{3.12}) in the form 
\be
f^{(2)}(\alpha, z) = \sum_{n=4}^{\infty}(\alpha^{2}-1)\, p_{n}(\alpha)\,z^{n}. 
\ee
Then, we have the following explicit polynomials $p_{n}(\alpha)$
{\small
\begin{align*}
p_{4} &=\frac{1}{120} \left(11 \alpha ^2+1\right), \\
p_{5} &= \frac{1}{60} \left(11
   \alpha ^2+1\right), \\
p_{6} &= \frac{-37 \alpha ^4+2609 \alpha ^2+236}{10080}, \\
p_{7} &= \frac{-37
   \alpha ^4+1069 \alpha ^2+96}{3360},\\
p_{8} &= \frac{251 \alpha ^6-42479 \alpha ^4+734269
   \alpha ^2+65399}{2016000},\\
p_{9} &= \frac{251 \alpha ^6-16579 \alpha ^4+201569 \alpha
   ^2+17799}{504000},\\
p_{10} &= \frac{-1569 \alpha ^8+479945 \alpha ^6-18258687 \alpha
   ^4+170711655 \alpha ^2+14943856}{399168000},\\
p_{11} &= \frac{-1569 \alpha ^8+181757 \alpha
   ^6-4716675 \alpha ^4+35862963 \alpha ^2+3112564}{79833600},\\
p_{12} &= \frac{1}{15256200960000}(1815004 \alpha
   ^{10}-873963991 \alpha ^8+56886151422 \alpha ^6-1106810435488 \alpha
   ^4\\
   &+7108850079574 \alpha ^2+611825905479),\\
p_{13} &= \frac{1}{2542700160000}(1815004 \alpha
   ^{10}-324264841 \alpha ^8+14101619632 \alpha ^6-218412314878 \alpha
   ^4  \numberthis \\
   &+1217781886564 \alpha ^2+103957984919),\\
 p_{14} &= \frac{1}{122049607680000}(-427265 \alpha
   ^{12}+297527391 \alpha ^{10}-29352540329 \alpha ^8+940279163283 \alpha
   ^6\\
   &-12079216519482 \alpha ^4+59671526161226 \alpha
   ^2+5053911319976),\\
p_{15} &= \frac{1}{3487131648000}(-85453 \alpha ^{12}+21753395 \alpha
   ^{10}-1412547837 \alpha ^8+35474720295 \alpha ^6\\
   &-389381682962 \alpha
   ^4+1731385519410 \alpha ^2+145526223952),\\
p_{16} &= \frac{1}{2240830797004800000}(226205741 \alpha
   ^{14}-215001129589 \alpha ^{12}+29836328448647 \alpha ^{10}\\
   &-1408641856774913
   \alpha ^8+28952603091655483 \alpha ^6-277672065926382407 \alpha
   ^4\\
   &+1125312507357000129 \alpha
   ^2+93890464607856909),\\
p_{17} &= \frac{1}{280103849625600000}(226205741 \alpha
   ^{14}-77720885089 \alpha ^{12}+7019243713307 \alpha ^{10}\\
   &-256841114655053
   \alpha ^8+4457607960596863 \alpha ^6-38008205857100927 \alpha
   ^4\\
   &+141820189774444089 \alpha ^2+11748922612161069).
\end{align*}
}

\section{Details of the derivation}
\label{app:details}

\subsection{The monodromy problem}

The results of Section \ref{sec:exact} are obtained by the monodromy method
\cite{Zamolodchikov:1985ie,Hartman:2013mia,Fitzpatrick:2014vua}. Let us briefly recall the 
main points. The function $f(\bm{\lambda},z)$ is related to the monodromy properties of the following 
equation specialised to the vacuum block case
\be
\label{B.1}
\psi''(w)+T(\bm{\lambda}, z; w)\,\psi(w)=0, 
\ee
where
\be
\label{B.2}
\begin{split}
T(\bm{\lambda}, z; w) &= \frac{6\,\lambda_{3}}{(1-w)^{2}}+6\,\lambda_{1}\,\bigg(
\frac{1}{w^{2}}+\frac{1}{(w-z)^{2}}+\frac{2}{w\,(1-w)}
\bigg)\\
&-c(\bm{\lambda}, z)\,\frac{z\,(1-z)}{w\,(w-z)\,(1-w)}.
\end{split}
\ee
The accessory parameter $c(\bm{\lambda}, z)$ is fixed by imposing a trivial monodromy along a 
contour encircling the two points $w=0, z$. Then, the semiclassical Virasoro block $f(\bm{\lambda}, z)$ is
obtained by integrating the relation \footnote{The second term in the l.h.s. of (\ref{B.3}) is due to a normalization 
factor $z^{2h_{1}}$ that we must take into account when matching the monodromy computation with our 
definition of the Virasoro block.}
\be
\label{B.3}
c(\bm{\lambda}; z) -\frac{12\,\lambda_{1}}{z}= \partial_{z}\,f(\bm{\lambda}, z),
\ee
with suitable boundary condition at $z\to 0$. The determination of the accessory parameter is a difficult 
analytical problem. \footnote{
Rigorous results that are somewhat related to our problem are presented 
in \cite{Menotti:2012wq,Menotti:2013bka,Menotti:2014kra} with applications in 
\cite{Kulaxizi:2014nma}.
} In our case, there is no known solution for generic $\bm{\lambda}$, but one can work out a perturbative expansion
in powers of $\lambda_{1}$ at generic $\lambda_{3}$. Plugging (\ref{3.3}) into (\ref{B.3}), and assuming that 
$\psi(w)$ admits a regular expansion around $\lambda_{1}=0$, we set 
\be
\label{B.4}
\begin{split}
\psi(w) &= \psi_{0}(w) + \lambda_{1}\, \psi_{1}(w)+ \lambda_{1}^{2}\, \psi_{2}(w)+\dots, \\
c(\bm{\lambda}, z) &= \lambda_{1}\, c_{1}(\lambda_{3}, z)+ \lambda_{1}^{2}\, 
c_{2}(\lambda_{3}, z)+\dots, \\
\end{split}
\ee
and solve the equations
\begin{subequations}
\begin{align}
\label{B.5a}
& \psi_{0}''(w) + T_{0}(\lambda_{3}; w)\,\psi_{0} = 0, \\
\label{B.5b}
& \psi_{1}''(w) + T_{0}(\lambda_{3}; w)\,\psi_{1} = -T_{1}(\lambda_{3}, z; w)\,\psi_{0}, \\
\label{B.5c}
& \psi_{2}''(w) + T_{0}(\lambda_{3}; w)\,\psi_{2} = -T_{2}(\lambda_{3}, z; w)\,\psi_{0}-T_{1}(\lambda_{3}, z; w)\,\psi_{1}, 
\end{align}
\end{subequations}
where
\be
\label{B.6}
\begin{split}
T_{0}(\lambda_{3}; w) &= \frac{6\,\lambda_{3}}{(1-w)^{2}}, \\
T_{1}(\lambda_{3}, z; w) &=6\,\bigg(\frac{1}{w^{2}}+\frac{1}{(w-z)^{2}}+\frac{2}{w\,(1-w)}\bigg)
-c_{1}(\lambda_{3}, z)\,\frac{z\,(1-z)}{w\,(w-z)\,(1-w)}, \\
T_{2}(\lambda_{3}, z; w) &= -c_{2}(\lambda_{3}, z)\,\frac{z\,(1-z)}{w\,(w-z)\,(1-w)}.
\end{split}
\ee
The solution at first order is well known and quite simple. One starts from the two independent solutions at 
leading order, {\em i.e.} (\ref{B.5a}). They  are
\be
\label{B.7}
\psi_{0}^{\pm}(w) = (1-w)^{\frac{1}{2}(1\pm\alpha)}, \qquad \alpha = \sqrt{1-24\,\lambda_3}.
\ee
Two linearly independent solutions to  (\ref{B.5b}) are then
\be
\label{B.8}
\psi_{1}^{\pm}(w) = \frac{1}{\alpha}\,\psi_{0}^{+} \,\int dw \, \psi_{0}^{-}\,T_{1}\,\psi_{0}^{\pm}
-\frac{1}{\alpha}\,\psi_{0}^{-}\,\int dw\, \psi_{0}^{+}\,T_{1}\,\psi_{0}^{\pm}.
\ee
The integrands in (\ref{B.8}) have only polar singularities around $w=0$ and $w=z$. Thus, $\psi_{1}^{\pm}$ have 
trivial monodromy around these two points when the sum of residues in $w=0,z$ vanishes. This gives immediately
(we identify again with little abuse of notation $c_{n}(\lambda_{3}(\alpha), z)
\to c_{n}(\alpha, z)$)
\be
\label{B.9}
c_{1}(\alpha,z) = 6\,\frac{\alpha-1+(1-z)^{\alpha}\,(1+\alpha)}{(1-z)(1-(1-z)^{\alpha})}.
\ee
Integrating (\ref{B.9})  with the condition $f^{(1)}(\lambda_{3}, z) = \mc O(z^{2})$ for $z\to 0$, one gets (\ref{3.4}).
In principle, one could solve (\ref{B.5c}) in the very same way, but $\psi_{1}^{\pm}$ will be integrated and it is rather difficult to control the analytic structure of the result. 
Indeed, some of the integrals in (\ref{B.8}) are definitely non trivial, like for 
example
\be
\label{B.10}
\begin{split}
\int dw\, [\psi_{0}^{+}(w)]^{2} & \,\bigg(\frac{1}{w^{2}}+\frac{1}{(w-z)^{2}}+\frac{2}{w\,(1-w)}\bigg) = \\
&= - (1-w)^{\alpha +1} \bigg(\frac{\, _2F_1\left(1,1,1-\alpha
   ;\frac{1}{w}\right)}{\alpha  w}-\alpha  \Gamma (\alpha +1) \,
   _2\tilde{F}_1(1,\alpha +1,\alpha +2;1-w)\\
   & +\frac{(w-1) \Gamma (-\alpha ) \,
   _2\tilde{F}_1\left(1,2,1-\alpha
   ;\frac{z-1}{z-w}\right)}{(w-z)^2}+\frac{1}{w}\bigg).
\end{split}
\ee
In the next subsection, we discuss what can be obtained at specialised values of the parameter
$\alpha$, see (\ref{B.7}), where the function $c_{2}(\alpha, z)$ is first expanded in powers of $z$ and then resummed, thus bypassing the above problems.

\subsection{Special cases}
\label{sec:specialized}

If we fix $\alpha=\sqrt{1-24\,\lambda_{3}}$ and work out the solution of (\ref{B.5c}) order by order around $z=0$, 
we see that $\psi_{1}^{\pm}$ in (\ref{B.8}) have only poles. The procedure leading to (\ref{B.9})
can then be repeated and it is possible to derive very long series expansions of the form 
\be
\label{B.11}
c_{2}(\alpha, z) = \sum_{n=0}^{\infty} c_{2,n}\,z^{n}.
\ee
We analysed these expansions for the special values 
\be
\label{B.12}
\alpha = \frac{1}{2}, \frac{1}{3}, \frac{1}{4}, \frac{1}{5}, \frac{1}{6},  \frac{1}{7},
\ee
and found that in all cases, the function $C(\alpha, t) = c_{2}(\alpha, 1-(1-t)^{\frac{1}{\alpha}})$
is the generating function of a holonomic sequence, {\em i.e.}
it obeys a differential equation of the form \cite{zeilberger1990holonomic}
\be
\label{B.13}
P(\alpha, t)\, \partial_{t}\, C(\alpha, t)+Q(\alpha, t)\, C(\alpha, t) + R(\alpha, t) = 0,
\ee
where $P, Q, R$ are polynomials in $t$. Once these polynomials are identified, the function $c_{2}(\alpha, z)$
may be determined by integrating (\ref{B.13}) and replacing $t=1-(1-z)^{\alpha}$. We identified the precise form of (\ref{B.13}) for the special values in (\ref{B.12}) by 
computing the coefficients $c_{2,n}$ in (\ref{B.11}) up to $n\sim 60$. Just to give an example, for $\alpha=\tfrac{1}{2}$, 
one finds the differential equation
\be
\label{B.14}
t\,(t-1)^{3}\,(t-2)^{2}\,\partial_{t}C(\tfrac{1}{2}, t)+(t-1)^{2}(t-2)\,(3t^{2}-10t+6)\,C(\tfrac{1}{2},t)-18\,t^{3}=0,
\ee
leading to 
\be
\label{B.15}
\begin{split}
c_{2}(\tfrac{1}{2}, z) & = -\frac{18 \left(z-6 \sqrt{1-z}-2\right)}{(z-1) z}\\
&-\frac{72
   \left(\left(\sqrt{1-z}+2\right) z-2 \left(\sqrt{1-z}+1\right)\right) \left(4
   \log \left(\sqrt{1-z}+1\right)-\log (16(1- z))\right)}{(z-1) z \left(z+2
   \sqrt{1-z}-2\right)}.
\end{split}
\ee
As a check, one can expand (\ref{B.15}) in powers of $z$ and integrate term by term recovering 
the expansion (\ref{3.12}) for $\alpha=\tfrac{1}{2}$. Of course, after some manipulation, one can also check 
{\em a posteriori} that 
(\ref{B.15}) is in agreement with (\ref{4.1}).
Unfortunately, inspection of $c_{2}(\alpha, z)$ for the values (\ref{B.12})
reveals that it is far from being trivial to identify a regularity. For instance, for $\alpha=\tfrac{1}{3}$, 
one finds by the same procedure the function 
\be
\label{B.16}
\begin{split}
c_{2}(\tfrac{1}{3}, z) & = \frac{24 \left(-z+5 (1-z)^{2/3}+5 \sqrt[3]{1-z}+2\right)}{(z-1) z}\\
&-\frac{144
   \left(\sqrt[3]{1-z}+1\right) \left(-z+(1-z)^{2/3}+\sqrt[3]{1-z}+1\right) \log
   \left(\frac{1}{3}
   \left(\sqrt[3]{1-z}+\frac{1}{\sqrt[3]{1-z}}+1\right)\right)}{\left(\sqrt[3]{1-z
   }-1\right)^2 (z-1) z},
\end{split}
\ee
and one may hope to guess a general formula. This works easily for the first lines of (\ref{B.15}) and (\ref{B.16}),
{\em i.e.} for the non-transcendental part of the result. Instead, the logarithmic parts are not that simple, despite their
compact form.
For instance, for $\alpha=\tfrac{1}{5}$ one has
\be
\label{B.17}
\begin{split}
& c_{2}(\tfrac{1}{5}, z)  =-\frac{72 \left(2 \sqrt[5]{1-z} z+7 z+5 \sqrt[5]{1-z}-5\right)}{5
   \left(\sqrt[5]{1-z}-1\right) (z-1) z}\\
   &+\frac{72 \left(\sqrt[5]{1-z}+1\right)}{5 \left(\sqrt[5]{1-z}-1\right)^3 (1-z)^{4/5}}
    \left[-5 \log
   \left((1-z)^{4/5}+(1-z)^{3/5}+(1-z)^{2/5}+\sqrt[5]{1-z}+1\right)\right.\\
   &\left. +2 \log (1-z)+5
   \log 5\right] +\frac{144 \left(\sqrt[5]{1-z}+1\right) \coth ^{-1}\left(\frac{\sqrt{5}
   \left((1-z)^{2/5}+1\right)}{\left(\sqrt[5]{1-z}-1\right)^2}\right)}{\sqrt{5}
   \left(\sqrt[5]{1-z}-1\right)^3 (1-z)^{4/5}},
\end{split}
\ee
and the appearance of the inverse hyperbolic cotangent becomes a major problem in the identification of some
regularity. The  
explicit results for $\alpha = \tfrac{1}{4}, \tfrac{1}{6}$ are collected for the reader's interest
in App.~\ref{app:spec} in simplified form in order to display their structure. 
Of course, they may be also recovered from (\ref{4.1}).

\bigskip
\noindent
Further explicit results may be obtained by the very same procedure 
for integer values 
\be
\label{B.18}
\alpha = 2, 3, 4, \dots.
\ee
Here, we do not care about the physical interpretation of these values and simply use them to explore 
the form of $c_{2}(\alpha, z)$ at more specific points. In all cases in (\ref{B.18}),  $c_{2}(\alpha, z)$
is simply a rational function. Explicit expressions are
\begin{subequations}
\begin{align}
\label{B.19a}
c_{2}(2,z) &= \frac{36 z^3}{(z-2)^3 (z-1)}, \\
\label{B.19b}
c_{2}(3,z) &= \frac{72 (z-2) z^3 \left(z^2-5 z+5\right)}{(z-1) \left(z^2-3 z+3\right)^3}, \\
\label{B.19c}
c_{2}(4,z) &= \frac{12 z^3 \left(9 z^6-98 z^5+466 z^4-1208 z^3+1784 z^2-1416
   z+472\right)}{(z-2)^3 (z-1) \left(z^2-2 z+2\right)^3}, \\
\label{B.19d}
c_{2}(5,z) &= \frac{12 (z-2) z^3}{(z-1) \left(z^4-5 z^3+10 z^2-10 z+5\right)^3} 
(12 z^8-147 z^7+826 z^6-2783 z^5 \notag \\
&+6104 z^4-8875 z^3+8325   z^2-4600 z+1150),
\end{align}
\end{subequations} 
and so on. Despite their relative simplicity, it is again non trivial to find a regularity in the sequence 
(\ref{B.19a}-\ref{B.19d}).

\subsection{Back to the monodromy problem}

We now come back to the second order monodromy problem in  (\ref{B.5c}) and try to exploit the 
many exact results of Section \ref{sec:specialized} in order to set up a mixed heuristic strategy.
Two independent solutions may be written as in (\ref{B.8}) and read
\be
\label{B.20}
\psi_{2}^{\pm}(w) = \frac{1}{\alpha}\,\psi_{0}^{+} \,\int dw \, \psi_{0}^{-}\,
(T_{1}\,\psi_{1}^{\pm}+T_{2}\,\psi_{0}^{\pm})
-\frac{1}{\alpha}\,\psi_{0}^{-}\,\int dw\, \psi_{0}^{+}\,
(T_{1}\,\psi_{1}^{\pm}+T_{2}\,\psi_{0}^{\pm}).
\ee
Expanding around $w=0$ or $w=z$, one finds 
\be
\label{B.21}
\begin{split}
\psi_{2}^{\pm}(w) &= \frac{1}{2}\,\log^{2}w+A^{\pm}(\alpha, z)\,\log w+\mc O(w\,\log w), \\
\psi_{2}^{\pm}(w) &= B^{\pm}(\alpha, z)\,\log^{2}(w-z)+C^{\pm}(\alpha, z)\,\log (w-z)+\mc O((w-z)\,\log(w-z)).
\end{split}
\ee
The monodromy cannot be read from these local expansions. Nevertheless, one observes that the integrand in the 
second term of (\ref{B.20}) for $\psi_{2}^{+}$ has a remarkable property. Its expansion around $w=0$ has the form 
\be
\label{B.22}
\psi_{0}^{+}\,
(T_{1}\,\psi_{1}^{+}+T_{2}\,\psi_{0}^{+}) = \frac{-\log w+f_{1}(\alpha, z)}{w^{2}}
+\frac{f_{2}(\alpha, z, c_{2})}{w}+\dots,
\ee 
with no $\log w$ in the residue of the simple pole. Also, the function $f_{2}(\alpha, z,c_{2})$ involves $c_{2}(\alpha, z)$
that is absent from the double pole. A similar property holds for the expansion around $w=z$
\be
\label{B.23}
\psi_{0}^{+}\,
(T_{1}\,\psi_{1}^{+}+T_{2}\,\psi_{0}^{+}) = \frac{-(1-z)^{\alpha+1}\,\log (w-z)+
\widetilde f_{1}(\alpha, z)}{(w-z)^{2}}
+\frac{\widetilde f_{2}(\alpha, z, c_{2})}{w-z}+\dots,
\ee 
Then, we  determine $c_{2}(\alpha, z)$ by the simple and well posed requirement that the sum of 
the simple pole residues in (\ref{B.22}) and (\ref{B.23}) vanishes, let us call it $\widehat c_{2}$
\be
\label{B.24}
f_{2}(\alpha, z, \widehat c_{2})+\widetilde f_{2}(\alpha, z, \widehat c_{2}) = 0.
\ee
Admittedly, 
this procedure has no particular rigorous motivation but just tries to get the most out of the expansions (\ref{B.22}) and 
(\ref{B.23}). In particular, there is no reason why $\widehat c_{2}$ has to be identified with the exact 
accessory function $c_{2}$ such that the exact monodromy is trivial. However, 
one can now compare this function $\widehat c_{2}$ with the exact solutions determined 
in Section \ref{sec:specialized} at the many special values (\ref{B.12}), (\ref{B.18}). Quite surprisingly,  the mismatch 
is always very simple
\be
\label{B.25}
c_{2}(\alpha, z) = \widehat c_{2}(\alpha, z) +18\,(\alpha+1)\frac{1+(1-z)^{\alpha}}{(1-z)\,[1-(1-z)^{\alpha}]}.
\ee
In other words, the term $ \widehat c_{2}(\alpha, z) $ captures all the transcendental contributions whose origin is from 
hypergeometric functions depending on $\alpha$. The remainder is the very simple function in the second term of (\ref{B.25}). Its origin is presumably from the contributions of the logarithms in the double pole of (\ref{B.22}) 
and (\ref{B.23}), although a global analysis of the exact $\psi_{2}^{\pm}$ would be necessary to clarify this.
The combination in the r.h.s. of (\ref{B.25}) is precisely the result in  (\ref{4.1}). It
 passes various non-trivial checks:
\begin{enumerate} 
\item It reproduces the exact result at the specialized points (\ref{B.12}) and (\ref{B.18}).
\item It agrees with the expansion (\ref{3.12}) at order $\mc O(z^{16})$ for generic $\alpha$. We recall that (\ref{3.12})
comes from the higher order solution of the Zamolodchikov recursion relations.
\item Finally, it is non perturbatively crossing symmetric, see (\ref{2.9}).
\end{enumerate}
Item (2) in the above list is perhaps the most stringent one because $\alpha$ is generic and the expansion (\ref{3.12})
-- including all terms in App.~\ref{app:f2} --
depends on it in a highly non-trivial way.
All these features lead us to conclude that (\ref{4.1}) is the correct expression for the next-to-leading 
semiclassical vacuum
block. In principle, it may be possible to make our derivation rigorous, but we feel that little doubts are left about the validity of (\ref{4.1}).

\section{Expressions of $c_{2}(\alpha, z)$ at $\alpha=\tfrac{1}{4},\tfrac{1}{6}$}
\label{app:spec}

We report here, for the reader's interest, the specialized function $c_{2}(\alpha, z)$ for 
$\alpha=\tfrac{1}{4},\tfrac{1}{6}$. They can be obtained from (\ref{4.1}), but we show them to display their
structure.
\be
\label{B.1}
\begin{split}
c_{2}(\tfrac{1}{4},z) &= \frac{9 \left(2 \left(7 (1-z)^{3/4}+7 \sqrt{1-z}+7 \sqrt[4]{1-z}+3\right)-3
   z\right)}{(z-1) z} \\
   & 
 -\frac{36}{\left(\sqrt[4]{1-z}-1\right)^2 (z-1) z} 
 \bigg[
 \left(2 \left(\sqrt[4]{1-z}+1\right)
   \left(\sqrt{1-z}+1\right)-\left(\sqrt[4]{1-z}+2\right) z\right)\\
   & \left(4 \log
   \left(\sqrt[4]{1-z}+1\right)+2 \log \left(\sqrt{1-z}+1\right)-\log (64(1-z))\right)
   \bigg]
   \end{split}
\ee
\be
\label{B.2}
\begin{split}
c_{2}(\tfrac{1}{6},z) &=-\frac{6 \left(12 \left(\sqrt[6]{1-z}-1\right)+\left(5 \sqrt[6]{1-z}+17\right)
   z\right)}{\left(\sqrt[6]{1-z}-1\right) (z-1) z}   	\\
   & 
   -\frac{24}{\left(\sqrt[6]{1-z}-1\right)^3 (1-z)^{5/6}}
   \bigg[ \left(\sqrt[6]{1-z}+1\right) \left(4 \log
   \left(\sqrt[6]{1-z}+1\right) \right. \\
   & \left. +\log \left(\sqrt[3]{1-z}-\sqrt[6]{1-z}+1\right)\right.\\
   &\left. +3
   \log \left(\sqrt[3]{1-z}+\sqrt[6]{1-z}+1\right)-\log (432
   (1-z))\right)\bigg]
   \end{split}
\ee

\section{Long expansion of the two-interval R\'enyi entropy}
\label{app:renyi}

Using (\ref{5.2}) and our extended data for the expansion (\ref{3.7}), we get
\be
S_{n, \rm vac} = \frac{c\,(n-1)(n+1)^{2}}{n^{3}}\,\sum_{k=2}^{\infty}\sigma_{k}\,z^{k}+\mc O(c^{0}),
\ee
with the explicit coefficients
{\small
\begin{align*}
\sigma_{2} &= -\frac{1}{144}, \\ 
\sigma_{3} &=  -\frac{1}{144}, \\ 
\sigma_{4} &= \frac{-1309 n^4+2 n^2+11}{207360
   n^4}, \\ 
\sigma_{5} &= \frac{-589 n^4+2 n^2+11}{103680 n^4}, \\ 
\sigma_{6} &= \frac{-805139 n^8+4244 n^6+23397 n^4+86
   n^2-188}{156764160 n^8}, \numberthis \\ 
\sigma_{7} &= \frac{-244439 n^8+1724 n^6+9537 n^4+86 n^2-188}{52254720
   n^8}, \\ 
\sigma_{8} &= \frac{-6459666587 n^{12}+56285106 n^{10}+312586347 n^8+4722748 n^6-10301973
   n^4-67854 n^2+58213}{1504935936000 n^{12}}, \\ 
\sigma_{9} &= \frac{-1491872987 n^{12}+15293106
   n^{10}+85282347 n^8+1833148 n^6-3985173 n^4-67854 n^2+58213}{376233984000
   n^{12}}, \\ 
\sigma_{10} &= \frac{1}{297977315328000 n^{16}}(-1098074352431 n^{16}+12818273224 n^{14}+71779112743
   n^{12}+2005530358 n^{10}\\
   &-4341928013 n^8-129198788 n^6+110407921 n^4+844006
   n^2-445820), \\ 
\sigma_{11} &= \frac{1}{59595463065600 n^{16}}(-205168198115 n^{16}+2665356736
   n^{14}+14986883227 n^{12}+514319494 n^{10}\\
   &-1108414169 n^8-48588236 n^6+41250877
   n^4+844006 n^2-445820), \\ 
\sigma_{12} &= \frac{1}{1366643159020339200000 n^{20}}(-4415649574925892347
   n^{20}+62766476909113526 n^{18}\\
   &+354351390797880981 n^{16}+14372653961026536
   n^{14}-30825885554247939 n^{12}-1808908135564014 n^{10}\\
   &+1523003046143151
   n^8+56147131454316 n^6-29483342794674 n^4-232322730364
   n^2+88390810828), \\ 
\sigma_{13} &= \frac{1}{227773859836723200000 n^{20}}(-693492488278292747
   n^{20}+10647680903036726 n^{18}\\
   &+60348261746649381 n^{16}+2814211318880136
   n^{14}-6006073680991539 n^{12}-443967283088814 n^{10}\\
   &+370233764078751
   n^8+20663431202316 n^6-10740178354674 n^4-232322730364
   n^2+88390810828), \\ 
\sigma_{14} &= \frac{1}{16399717908244070400000
   n^{24}}(-47220015602598891185
   n^{24}+775231178323314996 n^{22}\\
   &+4410547598118951299 n^{20}+231637262931595986
   n^{18}-491900587819492959 n^{16}-43930432559597964 n^{14}\\
   &+36253571634577911
   n^{12}+2776290861193866 n^{10}-1423709558722344 n^8-56824093196764
   n^6\\
   &+21453757559406 n^4+166480749880 n^2-49440922128), \\ 
\sigma_{15} &= \frac{1}{2342816844034867200000 n^{24}}(-6399731741802376253 n^{24}+111442004278820940
   n^{22}\\
   &+636367936312062263 n^{20}+37045104581062770 n^{18}-78277417467172875
   n^{16}-8194459617128580 n^{14}\\
   &+6688019137439955 n^{12}+659887041494970
   n^{10}-333028465921200 n^8-20581747259980 n^6\\
   &+7664791070238 n^4+166480749880
   n^2-49440922128), 
\end{align*}
}
and so on.

\bibliography{Virasoro-Biblio}
\bibliographystyle{JHEP}

\end{document}